\documentclass[aps,showpacs,twocolumn]{revtex4}
\usepackage{amsfonts}
\usepackage{amsmath}
\usepackage{graphicx}
\usepackage{epsfig}

\input{tcilatex}

\begin{document}

\title{On the effectiveness of the protocol creating the maximum
entanglement of two charge-phase qubits by cavity field}
\author{ C. Li}
\email{lichong@itp.ac.cn} \affiliation{Institute of Theoretical
Physics, The Chinese Academy of Science, Beijing, 100080, China}
\author{Y. B. Gao}
\affiliation{Institute of Theoretical Physics, The Chinese Academy
of Science, Beijing, 100080, China} \affiliation{Applied Physics
Department, Beijing University of Technology, Beijing, 100022,
China}
\date{January 12, 2004}

\begin{abstract}
We revisit the protocols to create maximally entangled states between two
Josephson junction (JJ) charge phase qubits coupled to a microwave field in
a cavity as a quantum data bus. We devote to analyze a novel mechanism of
quantum decoherence due to the adiabatic entanglement between qubits and the
data bus, the off-resonance microwave field. We show that even through the
variable of the data bus can be adiabatically eliminated, the entanglement
between the qubits and data bus remains and can decoher the superposition of
two-particle state. Fortunately we can construct a decoherence-free subspace
of two-dimension to against this adiabatic decoherence.To carry out the
analytic study for this decoherence problem, we develop Fr\H{o}hlich
transformation to re-derive the effective Hamiltonian of these system, which
is equivalent to that obtained from the adiabatic elimination approach .
\end{abstract}

\maketitle

\section{Introduction}

As a useful quantum resource, entanglement can not only be used to test
fundamental principles in quantum mechanics, such as Bell's inequalities,
but also play a central role in quantum information processing including
quantum computation, quantum teleportation and quantum cryptography.
Therefore how to create a stable and controllable entangled state in a
quantum bits (qubit) system is very important for quantum information
protocols \cite{qi}.

A number of protocols have been proposed to produce quantum entanglement in
different qubit systems, such as NMR, polarization photon, quantum dots,
Josephson junction. Due to the prompt progresses in preparing various solid
state qubits, these schemes become very promising to realize the practical
quantum computing. Actually, according to the DiVincenzo criteria the
couplings JJ qubits \cite{dd} for quantum computation, the solid system is
one of the best candidates for quantum computation, since qubit should be
scalable, controllable and with longer decoherence time. Actually it seems
difficult to fulfill all the requirements by quantum information processing.
Recently several groups have demonstrated the macroscopic quantum coherence
of Josephson junction (JJ) qubits with long decoherence time in experiments%
\cite{dd}\cite{y}\cite{3}\cite{4}\cite{4-2}\cite{4-3}.

Quantum entanglement plays the central role in integrating
multi-qubit to form a scalable quantum computing. We notice that,
in most of the protocols to produce such JJ qubit entanglement,
and correspondingly to carry out two qubit logic gate operations,
each qubit interacts with a common quantum object as a data bus,
which may be an electromagnetic cavity field, a quantum
transmission line coplanar cavity or an nano-mechanical resonator
\cite{han}\cite{5}. If the characterized frequency of the quantum
data bus is off-resonate to the energy spacing of the qubit, the
degree of freedom of the quantum data bus and the variables of the
quantum object can be separated adiabatically form that of two
qubit system. Then the induced inter-qubit interactions can create
an efficient quantum entanglement of two qubits.

However, as we have investigated\cite{sun1}, there usually exists quantum
entanglement between the states of data bus and those of the two qubit
system even after removing the data bus. This adiabatic quantum entanglement
has been studied according to the generalized Born-Oppenheimer (BO)
approximation\cite{sun1} where the slow variables can be driven by different
effective potentials provided by the fast internal states and then the
entanglement between fast and slow variables forms. Recently, Averin et al
\cite{Averin} similarly considered the adiabatic entanglement of two JJ
charge qubits. In this investigation, two JJ charge qubits are assumed to be
coupled with a large junction which works as a faster data bus. With the BO
adiabatic approximation, the energy of the lowest band of the latter
junction can be considered as the effective interaction between the two JJ
charge qubits. But the quantum decoherence induced by the adiabatic
entanglement has not been considered here though it may occur in the case
with higher excitation of large junction.

In this paper, we are also specific to the JJ qubit system coupling the
cavity and show that the adiabatic entanglement may cause the extra errors
of the logic gate operation for this two qubit system with high-excitation.
We will consider decoherence of the JJ qubit caused by the the thermal
excitation of large junction through this adiabatic entanglement mechanisim.
Actually, without considering thermal excitation, we are not clear if the
created entanglement between two JJ qubits are stable since it can be
produced according to an effective Hamiltonian, which is obtained in usual
by "ignoring" intermeddle variables of data bus \cite{pz}.

To carry out a totally analytic study, we utilize the generalized Fr\H{o}%
hlich transformation to re- derive the effective Hamiltonian of this system.
In this way we can study in details this novel decoherence phenomenon for
the entanglement of two JJ-qubits. There exist four entangled states for two
JJ qubit system, including two maximally entangled states that can be
obtained by controllable the micro-wave field. Though the superposition of
some two qubit states can decoher due to the adiabatic entanglement, there
exist a decoherence-free subspace, against to the decoherence induced by the
adiabatic separation process. Therefore, we found that only two of four
maximal entangled states are stable in this scheme.

\section{The model of two \ JJ\ qubits in cavity}

Without loss of generality, we investigate a simplified model,
which consisting of two JJ qubits in a cavity with a single mode
micro-wave fields (FIG. 1).
\begin{figure}[tbp]
\includegraphics[width=7cm]{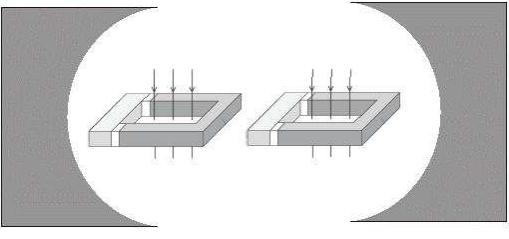}
\caption{ SQUID S1 and SQUID S2 in a cavity coupled to a microwave field. }
\label{fig:device}
\end{figure}
The Hamiltonian of the coupled system $H$ can be described as a
sum of that of the junctions, the cavity field \ and a interaction
term between the cavity and the junction \cite{han}\cite{5}, i.e,

\begin{eqnarray}
H &=&\hbar \omega a^{\dagger }a+4E_{C1}\left( n_{1}-n_{g1}\right)
^{2}-E_{J1}\left( \Phi \right) \cos \varphi _{1}  \notag \\
&&+4E_{C2}\left( n_{2}-n_{g2}\right) ^{2}-E_{J2}\left( \Phi \right) \cos
\varphi _{2},
\end{eqnarray}%
where
\begin{equation}
E_{C}=e^{2}/2\left( C_{g}+2C_{j}\right)
\end{equation}%
is the single-particle charging energy of the island, $C_{j}$ the
capacitance of the junction, $C_{g}$ the capacitance of gate and $\varphi
_{i}$ the phase difference between points on the opposite sides of the $i$%
-th junction. \ The Josephson coupling energy
\begin{equation}
E_{J}\left( \Phi \right) =2E_{J0}\cos \left( \frac{2\pi \Phi }{\Phi _{0}}%
\right)
\end{equation}%
depends on the the total flux $\Phi $ and the maximal coupling energy $%
E_{J0}=I_{c}\frac{\Phi _{0}}{2\pi }$. Here, $I_{c}$ is the critical current
of the junction, and $\Phi _{0}$ the total flux and flux quanta.

When a nonclassical microwave field with the vector potential%
\begin{equation}
\overrightarrow{A}\left( r\right) =\overrightarrow{u}_{\lambda }\left(
r\right) a+\overrightarrow{u}_{\lambda }^{\ast }\left( r\right) a^{\dagger }.
\end{equation}%
is applied, where $a^{\dagger }$ and $a$ are the creation and annihilation
operators of the cavity fields, the total flux $\Phi $ is divided into two
part
\begin{equation}
\Phi =\Phi _{e}+\Phi _{f}.
\end{equation}
$\Phi _{e}$ is static magnetic flux through the SQUIDs and
\begin{equation}
\Phi _{f}=\left\vert \Phi _{\mathrm{\lambda }}\right\vert \left( e^{-i\theta
}a+e^{i\theta }a^{\dagger }\right) ,
\end{equation}%
the microwave-filed-induced flux through the SQUIDs where
\begin{equation}
\Phi _{\mathrm{\lambda }}=\oint \overrightarrow{u}_{\lambda }\left( r\right)
\cdot d\overrightarrow{l}.
\end{equation}

We take $E_{C1}=E_{C2}$ and $E_{J1}\left( \Phi \right) =E_{J2}\left( \Phi
\right) $ and then the Hamiltonian of the system becomes%
\begin{eqnarray}
H &=&\varepsilon \left( V_{g}\right) \left( \sigma _{\mathrm{z}}^{\mathrm{1}%
}+\sigma _{\mathrm{z}}^{\mathrm{2}}\right) +\hbar \omega a^{\dagger }a
\notag  \label{MFAGrandCanonicalHamiltonian} \\
&-&2E_{J0}\cos \left( \dfrac{\pi \Phi _{e}+\Phi _{f}}{\Phi _{0}}\right)
\left( \sigma _{\mathrm{x}}^{\mathrm{1}}+\sigma _{\mathrm{x}}^{\mathrm{2}%
}\right) .
\end{eqnarray}%
where the quasi-spin operators $\sigma _{\mathrm{x}}$ , $\sigma _{\mathrm{y}%
},$ and $\sigma _{\mathrm{z}}$ are defined \ with respect to the the states $%
|0\rangle $ and $|1\rangle $ of no (one) excess cooper pair on the island.
To form a qubit or a two-level system, one need to tune the gate voltage $%
V_{g}$ so that $n_{g}$ is approximately a half-integer. In this case the
charge eigen-states are $|0\rangle $ and $|1\rangle $. We assume $\left\vert
\Phi _{\mathrm{\lambda }}\right\vert \ll \Phi _{0}$, and focus on the
charging regime $E_{C}\gg E_{J}$. Then, the Hamiltonian can be approximated
as

\begin{equation*}
H=H_{0}+H_{I}:
\end{equation*}

\begin{eqnarray}
H_{0} &=&\hbar \omega _{\mathrm{J}}\left( \sigma _{\mathrm{z}}^{\mathrm{1}%
}+\sigma _{\mathrm{z}}^{\mathrm{2}}\right) +\hbar \omega a^{\dagger }a
\notag \\
H_{I} &=&g\left( a+a^{\dagger }\right) \left( \sigma _{\mathrm{x}}^{\mathrm{1%
}}+\sigma _{\mathrm{x}}^{\mathrm{2}}\right)
\end{eqnarray}%
where%
\begin{equation*}
\hbar \omega _{\mathrm{J}}=2E_{C}\left[ \frac{C_{g}V_{g}}{e}-\left(
2n+1\right) \right]
\end{equation*}
and the coupling constant between qubit and the cavity field is

\begin{equation}
g=-2\dfrac{I_{c}\phi _{0}}{2\pi }\sqrt{\dfrac{h\nu }{2\mu _{0}}}\int_{S}%
\overrightarrow{e}\cdot d\overrightarrow{s}\sin \dfrac{\phi _{e}}{\phi _{0}}%
\pi .
\end{equation}

In practice we take the volume of the cavity and the wavelength of microwave
respectively as $\symbol{126}1\mathrm{cm}^{3}$ and \symbol{126}$1$\textrm{cm}%
, the dimension of the Josephson junction as $\symbol{126}1\mathrm{\mu m}$,
the critical current of the junction as $\ I_{\mathrm{c}}\symbol{126}10^{-5}%
\mathrm{A}$. Due to Eq$\left( \text{4}\right) $, we have
$\dfrac{g}{\hbar \omega }\ll 1$, which means $H_{I}\ll H_{0}$. So
we can perform perturbation theory represented by a generalized
Fr\H{o}hlich transformation\cite{sun} on the Hamiltonian$\left(
\mathrm{2}\right) $. Then we can obtain the effective Hamiltonian
of two JJ qubit by removing the variables of the microwave field
approximately.

\section{The effective Hamiltonian from the generalized Fr\H{o}lich
transformation}

In its original approach for superconductivity BCS theory, the Fr\H{o}hlich
transformation\cite{sun} is utilized to get the effective Hamiltonian for
electron-electron interaction from electron-phonon interaction. In general
we can consider a interaction system described by a sum of free Hamiltonian
and interaction Hamiltonian,

\begin{equation}
H=H_{0}+H_{I}\text{ }.
\end{equation}%
Comparing with the free part $H_{0},$the interaction part $H_{I}$
can be \ regard as a perturbation. Let us define an anti-Hermitian
operator $S$, and a corresponding unitary operator $U^{\dag }=\exp
\{-S\}$. We perform an unitary transformation on the
Hamiltonian$\left( \text{11}\right) $ by this unitary operator,
and then get the equivalent Hamiltonian \ as

\begin{eqnarray}
H &=&U^{\dagger }HU  \notag  \label{MFAGrandCanonicalHamiltonian} \\
&=&H_{0}+\underset{\mathrm{n=1}}{\sum }\frac{\left( -1\right) ^{n}}{(n+1)!}%
\underset{n}{[\underbrace{S,[\cdots \lbrack S,[S,}}H_{I}]]\cdots ]].
\end{eqnarray}

Since the unitary transformation $U$ is time -independent, Hamiltonians $%
\left( \text{11}\right) $ and $\left( \text{12}\right) $ describe
the same physical process. We can take both the interaction
$H_{I}$ and operator $S$ in the first order terms in the right
hand side. At the same time, we require the operator $S$ \ to
satisfy the following condition

\begin{equation}
H_{\mathrm{I}}+[H_{\mathrm{0}},S]=0.
\end{equation}%
In Eq.$\left( \text{12}\right) $, if we discard the higher order
terms and only keep the second-order term, the effective
Hamiltonian can be achieved approximately as

\begin{equation}
H_{\mathrm{eff}}\cong H_{\mathrm{0}}+\frac{1}{2}\left[ H_{\mathrm{I}},S%
\right] .
\end{equation}%
From the Eq.$\left( \text{13}\right) $ we certainly know how to
construct the anti-Hermitian operator $S$, which has the following
form \

\begin{equation}
S=\underset{\mathrm{m}\neq \mathrm{n}}{\sum }\frac{\left( H_{\mathrm{I}%
}\right) _{\mathrm{mn}}}{E_{\mathrm{m}}-E_{\mathrm{n}}}\left\vert
m\right\rangle \left\langle n\right\vert ,\text{ \ }
\end{equation}%
where $\left\vert m\right\rangle $ and $E_{m}$ are the eigenvectors and
eigenvalues of $H_{0}$ respectively. The transformation, by which one can
draw out effective Hamiltonian$\left( \text{14}\right) $ from the Hamiltonian$%
\left( \text{11}\right) $, is the so-called general Fr\H{o}hlich
transformation. It has been proved in Ref\cite{sun} that this generalized Fr%
\H{o}hlich transformation is just equivalent to the second-order
perturbative theory.

Now we use the above approximation method to derive the effective
Hamiltonian for the two-JJ qubit entanglement. Under the condition
$g$ $\ll $ $\hbar \omega $, we explicitly construct the
anti-Hermitian operator $S$ of Eq.$\left( \text{3}\right) $ as
following

\begin{eqnarray}
S &=&\dfrac{g}{2}\{\Delta _{+}\left( a-a^{\dagger }\right) \left( \sigma _{%
\mathrm{x}}^{\mathrm{1}}+\sigma _{\mathrm{x}}^{\mathrm{2}}\right)  \notag
\label{criticaltemperature} \\
&&+i\Delta _{-}\left( a+a^{\dagger }\right) \left( \sigma _{\mathrm{y}}^{%
\mathrm{1}}+\sigma _{\mathrm{y}}^{\mathrm{2}}\right) \}
\end{eqnarray}%
where the coefficients
\begin{equation}
\Delta _{\pm }=\left( \dfrac{1}{\hbar \omega -2\hbar \omega _{\mathrm{J}}}%
\pm \dfrac{1}{\hbar \omega +2\hbar \omega _{\mathrm{J}}}\right)
\end{equation}

Using the above explicit expression for anti-Hermitian operator $S$, we can
finish the generalized Fr\H{o}hlich transformation and then obtain effective
Hamiltonian obviously.\bigskip
\begin{eqnarray}
H_{\mathrm{eff}} &=&\hbar \omega _{\mathrm{J}}\left( \sigma _{\mathrm{z}}^{%
\mathrm{1}}+\sigma _{\mathrm{z}}^{\mathrm{2}}\right) -\dfrac{g^{2}}{2}\Delta
_{-}\left( a+a^{\dagger }\right) ^{2}\left( \sigma _{z}^{1}+\sigma
_{z}^{2}\right)  \notag  \label{criticaltemperature} \\
&&+\hbar \omega a^{\dagger }a-\dfrac{g^{2}}{2}\Delta _{+}-\dfrac{g^{2}}{2}%
\Delta _{+}\sigma _{\mathrm{x}}^{\mathrm{1}}\sigma _{\mathrm{x}}^{\mathrm{2}%
}.
\end{eqnarray}%
If the micro-wave field is very weak, we can discard the second terms of $%
a^{2}$ and $a^{\dagger 2}$ in the effective Hamiltonian under the rotating
wave approximation. Then the effective Hamiltonian Eq.$\left( \mathrm{18}%
\right) $ reads

\begin{eqnarray}
H_{\mathrm{eff}} &=&\left( \hbar \omega _{\mathrm{J}}+g^{2}\Delta
_{-}a^{\dagger }a\right) \left( \sigma _{\mathrm{z}}^{\mathrm{1}}+\sigma _{%
\mathrm{z}}^{\mathrm{2}}\right)  \notag \\
&&+\hbar \omega a^{\dagger }a+\dfrac{g^{2}}{2}\Delta _{+}\sigma _{\mathrm{x}%
}^{\mathrm{1}}\sigma _{\mathrm{x}}^{\mathrm{2}}.
\end{eqnarray}%
or

\begin{eqnarray}
H_{\mathrm{eff}} &=&\sum_{n}H(n)|n\rangle \langle n|:  \notag \\
H(n) &=&\left( \hbar \omega _{\mathrm{J}}+ng^{2}\Delta _{-}\right) \left(
\sigma _{\mathrm{z}}^{\mathrm{1}}+\sigma _{\mathrm{z}}^{\mathrm{2}}\right) \\
&&+\hbar \omega n+\dfrac{g^{2}}{2}\Delta _{+}\sigma _{\mathrm{x}}^{\mathrm{1}%
}\sigma _{\mathrm{x}}^{\mathrm{2}}.  \notag
\end{eqnarray}

In general this is a typical effective Hamiltonian leading \ the two-qubit
quantum logic gate. In usual it is obtained by adiabatically eliminating the
variable data bus with various methods\cite{pz}. However, in most of
previous works, this crucial terms
\begin{equation}
g^{2}\Delta _{-}a^{\dagger }a\left( \sigma _{\mathrm{z}}^{\mathrm{1}}+\sigma
_{\mathrm{z}}^{\mathrm{2}}\right)
\end{equation}%
referred to the ac Stark effect (a dispersive frequency shift effects) has
been either irrationally ignored or passed over in silence. This is
unsatisfactory even though we can now prove that it can run the logic gate
of two qubit system in next section.

\section{A novel decoherence mechanism: the inverse Stern-Gerlach effect}

Having gotten the effective Hamiltonian with a perfect inter-qubit
interaction, we can show how to create the quantum entanglement by
controllable the coupling between photon and qubit. As a data bus, the role
of cavity field is to introduce extra controllable parameters. In the next
section we will show the details to implemental an ideal two qubit logical
gate operations in the decoherence free subspace (DFS)\cite{97c}. However,
for those states outside the DFS, we can demonstrate a novel decoherence
phenomenon related to the so-called inverse Stern-Gerlach\cite{inverse}
effect from the adiabatic variable separation based on the BO approximation
\cite{sun}.

Let us generally consider the adiabatic evolution of two identical charge
qubits $1$ and $2$, coupled to a single-mode field in the microwave cavity.
In the off-resonance case, the motion of the qubits does not excite the
transitions from a cavity mode to another, and then the photon number is
conserved, i.e.,
\begin{equation}
\lbrack H_{\mathrm{eff}},a^{\dagger }a]=0
\end{equation}%
This shows that the total wave function will adiabatically keep the
factorized structure
\begin{equation}
\left\vert \Psi (t)\right\rangle =\left\vert \phi _{n}(t)\right\rangle
\otimes \left\vert n\right\rangle
\end{equation}%
during evolution only if the cavity is exactly prepared initially in a
single number state with definite phonon number, namely, a Fock state $%
\left\vert n\right\rangle .$ In this case the qubit part is just governed by
the effective Hamiltonian $\langle n|H_{\mathrm{eff}}\left\vert
n\right\rangle $ $=H(n)$ and then we can manipulate the qubit system
according to the $n$-dependent Hamiltonian to form maximal entanglement.
However, if one can not prepare the cavity in a single Fock state, the part $%
\left\vert \phi (t)\right\rangle $ of qubit must depend on the different
phonon number $n$ and then we can not make an exact manipulation for qubit
part due to this correlation to cavity field. This kind feature of quantum
adiabatic entanglement is just a novel physical source of the quantum
decoherence in the process of two qubit logical gate operations.

We remark that this phenomenon is an analog of "inverse Stern-Gerlach
effect" in atomic optics, in which discrete atomic trajectories are
correlated to different photon numbers in the cavity. When the atom is
non-resonant with the cavity modes, there appears a dispersive frequency
shift effects affecting both the atomic transition and the field mode. It
can be interpreted as single atom and single photon index effects. These
effects lead to various interesting potential applications, which have been
investigated in cavity QED\ for atoms, both theoretically and
experimentally, e.g., the interference schemes to measure matter-wave phase
shifts produced by the non resonant interaction \cite{inverse}. This schemes
performs a quantum non-demolition measurement of photon numbers in a cavity,
at the single photon level. Its experimental demonstration is based on the
detection of Ramsey resonances on circular Rydberg atoms crossing a very
high Q cavity. For this kind of "inverse Stern Gerlach effect", we even
presented an extensive generalization based on the Born-Openheimer
approximation to analyzed the the adiabatic separation induced quantum
entanglements \cite{sun}. Thus it defines the adiabatic quantum decoherence
in general case. We can discuss this effect for a solid state based system
with two charge qubit.

To see the quantum decoherence due to the generalized "inverse Stern Gerlach
effect", we assume the two JJ qubits and the cavity field are initially
prepared in a factorizable state:
\begin{equation}
\left\vert \Psi \left( 0\right) \right\rangle =\left\vert \phi \right\rangle
\otimes \left\vert \varphi \right\rangle
\end{equation}%
where $\left\vert \phi \right\rangle $ is the initial state of the two JJ\
qubits and $\left\vert \varphi \right\rangle $ the state of the field. In
general, if the cavity is prepared initially in a superposition state of
Fock state
\begin{equation}
\left\vert \varphi \right\rangle =\sum_{n}c_{n}\left\vert n\right\rangle
\end{equation}%
rather than a single Fock states, the total system will evolve according to
\begin{equation}
\left\vert \Psi (t)\right\rangle =\sum_{n}c_{n}\left\vert \phi
_{n}(t)\right\rangle \otimes \left\vert n\right\rangle
\end{equation}%
where
\begin{equation}
\left\vert \phi _{n}(t)\right\rangle =U_{n}(t)\left\vert \phi
(0)\right\rangle .
\end{equation}%
The effective evolution matrix $U_{n}(t)=\exp [-iH(n)t]$ \ is governed by $%
H(n).$This "inverse Stern Gerlach effect" result from the dependence $%
\left\vert \phi _{n}\left( t\right) \right\rangle $ to different $n$.

Let $\left\vert m\right\rangle $ be the single Fock state that we want to
prepare and $H(m)$ be the controlled Hamiltonian. Then we can characterize
the difference between the real evolution and the ideal one $\rho _{m}\left(
t\right) =\left\vert \phi _{m}(t)\right\rangle \langle \phi _{m}(t)|,$ by
the fidelity
\begin{equation}
F=Tr[\rho _{m}\left( t\right) \rho \left( t\right)
]=\sum_{n}|c_{n}|^{2}\left\vert \langle \phi _{m}(t)|\phi
_{n}(t)\right\rangle |^{2}
\end{equation}%
where
\begin{equation}
\rho \left( t\right) =Tr_{C}\left( \left\vert \Psi (t)\right\rangle \langle
\Psi (t)|\right) =\sum_{n}|c_{n}|^{2}\left\vert \phi _{n}(t)\right\rangle
\langle \phi _{n}(t)|
\end{equation}%
is the reduced density matrix of the two JJ qubits.

Usually it is difficult to prepare the Fock state $\left\vert m\right\rangle
$ and we can only use the coherent state $\left\vert \alpha \right\rangle $
with average photon number $\langle \alpha |a^{\dagger }a\left\vert \alpha
\right\rangle $ $=$ $m.$ Then we assume the junctions are initially in the
state $\left\vert 0\right\rangle _{1}\left\vert 0\right\rangle _{2}$ , and
the initial state of the micro-wave field is the coherent state $\left\vert
\alpha \right\rangle $. Then total system will evolve into \

\begin{eqnarray}
\left\vert \Psi \right\rangle &=&e^{-\frac{1}{2}\left\vert m\right\vert ^{2}}%
\underset{n}{\sum }\dfrac{m^{n}}{n!}\left\vert \phi
_{n}^{00}(t)\right\rangle \left\vert n\right\rangle :  \notag \\
\left\vert \phi _{n}^{00}(t)\right\rangle &=&\{c_{2}^{\ast }\left( n\right)
\left\vert 0\right\rangle _{1}\left\vert 0\right\rangle _{2}+ic_{1}\left(
n\right) \left\vert 1\right\rangle _{1}\left\vert 1\right\rangle _{2}\},
\end{eqnarray}%
where the time-dependent coefficients are

\bigskip
\begin{equation*}
\left\{
\begin{array}{c}
c_{1}\left( n\right) =\sin \left( \omega _{n}t\right) \cos \left( \theta
_{n}\right) , \\
c_{2}\left( n\right) =\cos \left( \omega _{n}t\right) -i\sin \left( \omega
_{n}t\right) \sin \left( \theta _{n}\right) , \\
\omega _{n}=\frac{1}{\hbar }\sqrt{\left( 2\hbar \omega _{\mathrm{J}%
}+2g^{2}\Delta _{-}n\right) ^{2}+(\dfrac{g^{2}}{2}\Delta _{+})^{2}}, \\
\sin \left( \theta _{n}\right) =\dfrac{2\hbar \omega _{\mathrm{J}%
}+2g^{2}\Delta _{-}n}{\sqrt{\left( 2\hbar \omega _{\mathrm{J}}+2g^{2}\Delta
_{-}n\right) ^{2}+(\dfrac{g^{2}}{2}\Delta _{+})^{2}}}.%
\end{array}%
\right.
\end{equation*}

Through a simple calculation, we obtain the fidelity

\begin{eqnarray}
&&F=e^{-\frac{1}{2}\left\vert m\right\vert ^{2}}\underset{n}{\sum }\dfrac{%
m^{n}}{n!}\{\cos ^{2}\left( \omega _{n}t\right) \cos ^{2}\left( \omega
_{m}t\right)  \notag \\
&+&\sin ^{2}\left( \omega _{n}t\right) \sin ^{2}\left( \theta _{n}\right)
\sin ^{2}\left( \omega _{m}t\right) \sin ^{2}\left( \theta _{m}\right)
\notag \\
&+&\cos ^{2}\left( \omega _{m}t\right) \sin ^{2}\left( \omega _{n}t\right)
\sin ^{2}\left( \theta _{n}\right)  \notag \\
&+&\cos ^{2}\left( \omega _{n}t\right) \sin ^{2}\left( \omega _{m}t\right)
\sin ^{2}\left( \theta _{m}\right) \\
&+&\sin ^{2}\left( \omega _{n}t\right) \cos ^{2}\left( \theta _{n}\right)
\sin ^{2}\left( \omega _{m}t\right) \cos ^{2}\left( \theta _{m}\right)
\notag \\
&+&2\sin \left( \omega _{n}t\right) \cos \left( \omega _{n}t\right) \cos
\left( \theta _{n}\right) \sin \left( \omega _{m}t\right) \cos \left( \theta
_{m}\right) \cos \left( \omega _{m}t\right)  \notag \\
&+&2\sin ^{2}\left( \omega _{n}t\right) \sin ^{2}\left( \omega _{m}t\right)
\cos \left( \theta _{n}\right) \cos \left( \theta _{m}\right) \sin \left(
\theta _{m}\right) \sin \left( \theta _{n}\right) \}.  \notag
\end{eqnarray}

FIG.2 and FIG.3 illustrate that the fidelity decays sharply with the value
of $m$, namely, coherent state $\left\vert \alpha \right\rangle $ leads to
big deviation of $\rho _{\alpha }$ and $\rho _{m}$ with big $\alpha $. This
is because that coherent state $\left\vert \alpha \right\rangle $ is just in
Fock state $\left\vert m\right\rangle $ with the probability $P_{m}$,

\begin{equation*}
\left. P_{m}\right\vert _{\alpha =m}=\left. \left\vert \left\langle m\right.
\left\vert \alpha \right\rangle \right\vert ^{2}\right\vert _{\alpha
=m}=e^{-\left\vert m\right\vert ^{2}}\frac{\left\vert m\right\vert ^{2m}}{m!}%
.
\end{equation*}%
When $m$ $\rightarrow \infty $, $P_{m}\rightarrow 0$ quickly.

\begin{figure}[tbp]
\includegraphics[angle=-90,width=7cm]{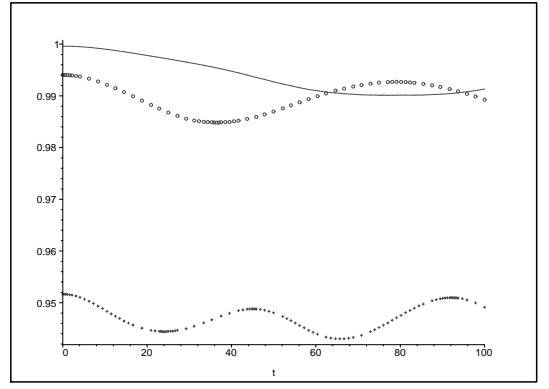}
\caption{ Fidelity $F$ as a function of time $t$, with different $m$, which
is the mean eigenvalue of the micro-wave field. $m=0.2$(line), $m=0.4$%
(circles) and $m=0.7$ (crosses).}
\label{fig:time}
\end{figure}
\begin{figure}[tbp]
\includegraphics[angle=-90,width=7cm]{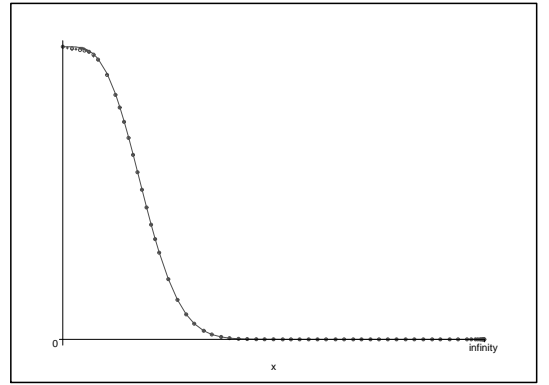}
\caption{ Fidelity $F$ as a function of the mean eigenvalue of the
micro-wave field $m$, with different fixed time $t=13$(line),$t=40$(circles)
and $t=70$ (crosses). }
\label{fig: mean}
\end{figure}

The above discussion shows us that, when we prepare the controllable cavity
field in different initial states, one can get different entangled states
for the two qubit. This motivates us to explore the possibility to realize
the perfect logic gate operation by initially preparing the micro-wave field
in coherent state and Fock state. Let us consider the above mentioned
problem in the following.

We aim to get a standard Bell state $\left\vert \phi ^{+}\right\rangle $
\begin{equation*}
\left\vert \phi ^{+}\right\rangle =\frac{1}{\sqrt{2}}\left( \left\vert
00\right\rangle +\left\vert 11\right\rangle \right)
\end{equation*}
from the state $\left\vert 0\right\rangle _{1}\left\vert 0\right\rangle _{2}$%
, by preparing the cavity field initially in Fock state $\left\vert
k\right\rangle $. The evolution from $\left\vert 0\right\rangle
_{1}\left\vert 0\right\rangle _{2}$ to $\left\vert \phi ^{+}\right\rangle $\
naturally realize a perfect ideal logic gate.

The real evolution process governed by the effective adiabatic Hamiltonian $%
H(k)$ is
\begin{eqnarray*}
\left\vert \phi _{k}^{00}(t)\right\rangle &=&e^{-i\frac{H\left( k\right) }{%
\hbar }t}\left\vert 00\right\rangle \\
&=&c_{2}^{\ast }\left( k\right) \left\vert 00\right\rangle +ic_{1}\left(
k\right) \left\vert 11\right\rangle \},
\end{eqnarray*}%
where the Hamiltonian $H(k)$ corresponds to the Fock state $\left\vert
k\right\rangle $ for fixed $k$. We can use the square of the norm of the
inner product
\begin{equation*}
f_{k}=\left\vert \left\langle \phi _{k}^{00}(t)\right\vert \phi ^{+}\rangle
\right\vert ^{2}
\end{equation*}
to characterize the difference between the ideal state $\left\vert \phi
^{+}\right\rangle $\ and the real state $\left\vert \phi
_{k}^{00}\right\rangle $
\begin{eqnarray}
&&f_{m}=\left\vert \left\langle \phi _{m}^{00}(t)\right\vert \phi
^{+}\rangle \right\vert ^{2} \\
&=&\frac{1}{2}\left\vert \cos ^{2}\left( \omega _{m}t\right) +\sin
^{2}\left( \omega _{m}t\right) \left[ \cos \left( \theta _{m}\right) +\sin
\left( \theta _{m}\right) \right] ^{2}\right\vert .  \notag
\end{eqnarray}
\begin{figure}[tbp]
\includegraphics[angle=-90,width=7cm]{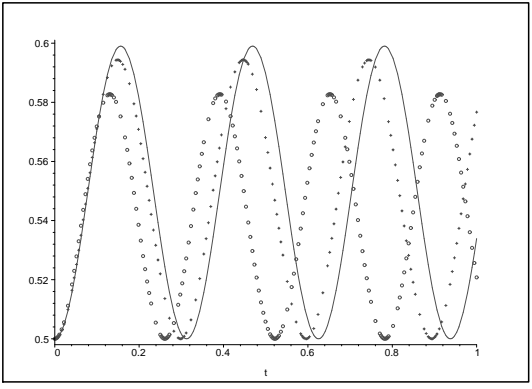}
\caption{The vertical axis represent the function $f_{m}$, the horizonal
axis represent time $t$, $m$ is the eigenvalue of the Fock state, $m =0$
(line), $m =10$(crosses) and $m =20$ (circles)}
\label{fig:bell-k}
\end{figure}
The above equation shows that $f_{m}$ is a periodic function of time $t$.

FIG.4 shows that $f_{m}$ decays with the average photon number $m$ and the
maximum value of $f_{m}$ can not approach 1. Based on this result, we can
not construct an ideal logic gate in this system.

Now we consider another case that the micro-wave field is prepared in
coherent state $\left\vert \alpha \right\rangle $ initially, and the
junctions is prepared initially in the state $\left\vert 0\right\rangle
_{1}\left\vert 0\right\rangle _{2}$. By a simple calculation, we get the
final state depicted by the reduced density matrix(RDM)

\begin{eqnarray}
\rho _{\alpha } &=&tr_{mw}\left\{ \left\vert \Psi \right\rangle \left\langle
\Psi \right\vert \right\}  \notag \\
&=&e^{-\left\vert \alpha \right\vert ^{2}}\underset{k}{\sum }\dfrac{%
\left\vert \alpha \right\vert ^{2k}}{k!k!}\left\vert \psi _{k}\right\rangle
\left\langle \psi _{k}\right\vert .
\end{eqnarray}%
We explicitly calculate the function
\begin{eqnarray}
&&f_{\alpha }=\left\vert \left\langle \phi ^{+}\right\vert \rho _{\alpha
}\left\vert \phi ^{+}\right\rangle \right\vert  \notag \\
&=&\frac{1}{2}e^{-\left\vert \alpha \right\vert ^{2}}\underset{k}{\sum }%
\dfrac{\left\vert \alpha \right\vert ^{2k}}{k!k!}|\cos ^{2}\left( \omega
_{k}t\right)  \notag \\
&+&\sin ^{2}\left( \omega _{k}t\right) \left[ \cos \left( \theta _{k}\right)
+\sin \left( \theta _{k}\right) \right] ^{2}|
\end{eqnarray}

\begin{figure}[tbp]
\includegraphics[angle=-90,width=7cm]{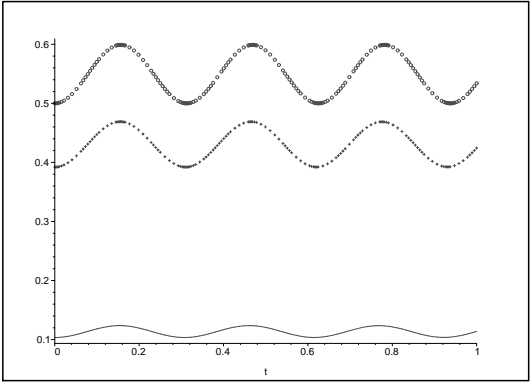}
\caption{ The vertical axis represent the function $f_{\protect\alpha }$,
the horizonal axis represent time $t$, and $\protect\alpha $ is the
eigenvalue of the coherent state. $\protect\alpha =0.1$ (circles),$\protect%
\alpha =1.1$ (crosses) and $\protect\alpha =5$ (line) }
\label{fig:bell-cohen}
\end{figure}

FIG.5 displays the evolution of the function $f_{\alpha }$ calculated from
eq.(34) for the qubits with coherent microwave fields. When the eigenvalue $%
\alpha \rightarrow \infty $, the fidelity $F\rightarrow 0$ .

It is shown from the FIG.4 and FIG.5, whatever the state of the microwave
field is prepared in, the ideal logic gate operation can not be\ realized in
this system. But this does not means that we can not obtain any maximal
entangled state in this way. We will discuss this problem in next section.

\bigskip

\section{Creating Maximal Entanglement in the Decoherence Free Subspace}

From the discussions in the above section, we find that the maximally
entangled state can not be obtained only from the state $\left\vert
0\right\rangle _{1}\left\vert 0\right\rangle _{2}$ and $\left\vert
1\right\rangle _{1}\left\vert 1\right\rangle _{2}$, even though one can
prepared external controlled microwave cavity field in an arbitary state.
While the other two states $\left\vert 0\right\rangle _{1}\left\vert
1\right\rangle _{2}$ and $\left\vert 1\right\rangle _{1}\left\vert
0\right\rangle _{2}$ can span a decoherence-free subspace(DFS)$\mathfrak{W}%
^{1}$, it means that any superposition of the state $\left\vert
0\right\rangle _{1}\left\vert 1\right\rangle _{2}$ and $\left\vert
1\right\rangle _{1}\left\vert 0\right\rangle _{2}$ can evolve into
this kind of DFS. Easily seen in Eq.(19), the effective
interaction between the
cavity and qubits $g^{2}\Delta _{-}a^{\dagger }a\left( \sigma _{\mathrm{z}}^{%
\mathrm{1}}+\sigma _{\mathrm{z}}^{\mathrm{2}}\right) $ vanishes in the DFS
and can not distinguish between any two states in this DFS. So we conclude
that there is not a "which-way detection" to determine the "paths" in this
case, i.e., there is not decoherence appearing in DFS.

Let us use a special example to demostrate the above observation. When
junctions are prepared initially in the state $\left\vert 0\right\rangle
_{1}\left\vert 1\right\rangle _{2}$ and the manipulative field prepared in
the Fock state $\left\vert n\right\rangle $, then the evolution of the total
system will evolve into

\begin{eqnarray}
\left\vert \psi _{01}\right\rangle \left\vert n\right\rangle
&=&e^{-iHt}\left\vert 0\right\rangle _{1}\left\vert 1\right\rangle
_{2}\left\vert n\right\rangle  \notag \\
&=&\cos \left( g^{2}\Delta _{-}t\right) \left\vert 0\right\rangle
_{1}\left\vert 1\right\rangle _{2}\left\vert n\right\rangle  \notag \\
&&-i\sin \left( g^{2}\Delta _{-}t\right) \left\vert 1\right\rangle
_{1}\left\vert 0\right\rangle _{2}\left\vert n\right\rangle
\end{eqnarray}%
It is obvious that, when $t=\tfrac{\pi \hbar \left( \omega ^{2}-4\omega _{%
\mathrm{J}}^{2}\right) }{16g^{2}\omega _{J}},$the two qubit system reaches a
maximally entangled state
\begin{equation}
\left\vert \psi _{01}\right\rangle =\dfrac{1}{\sqrt{2}}\left( \left\vert
0\right\rangle _{1}\left\vert 1\right\rangle _{2}-i\left\vert 1\right\rangle
_{1}\left\vert 0\right\rangle _{2}\right) .
\end{equation}%
By the same way, we can obtain the other maximal entangled state

\begin{equation}
\left\vert \psi _{10}\right\rangle \dfrac{1}{\sqrt{2}}\left( \left\vert
0\right\rangle _{1}\left\vert 1\right\rangle _{2}+i\left\vert 1\right\rangle
_{1}\left\vert 0\right\rangle _{2}\right) ,
\end{equation}%
when we take the $t=\tfrac{3\pi \hbar \left( \omega ^{2}-4\omega _{\mathrm{J}%
}^{2}\right) }{16g^{2}\omega _{J}}$. Both $\left\vert \psi
_{01}\right\rangle $ and $\left\vert \psi _{10}\right\rangle $ are
independent of the controllable micro-wave field, they belong to the DFS $%
\mathfrak{W}^{1}$. The basis $\left\vert 0\right\rangle _{1}\left\vert
0\right\rangle _{2}$ and $\left\vert 1\right\rangle _{1}\left\vert
1\right\rangle _{2}$ span the other subspace of the Hilbert space of the JJ
qubits($\mathfrak{W}$), we denote \bigskip $\mathfrak{W}_{\perp }^{1}$. Thus
we have $\mathfrak{W=W}^{1}\mathfrak{\oplus W}_{\perp }^{1}$.\bigskip\

\section{Overall Quality of\ Created Entanglements}

In above section we have discussed that when junctions are prepared
initially in the state $\left\vert 0\right\rangle _{1}\left\vert
0\right\rangle _{2}$ or state $\left\vert 1\right\rangle _{1}\left\vert
1\right\rangle _{2}$, we can not obtain maximally entangled state of the
junctions with any controllable microwave field. But we can study the
entanglement of these states evolved from the initial state $\left\vert
0\right\rangle _{1}\left\vert 0\right\rangle _{2}$ or state $\left\vert
1\right\rangle _{1}\left\vert 1\right\rangle _{2}$.

As well-known, for a bipartite system, composing of two subsystems $A$ and $%
B $, the bipartite entanglement can be measured by its concurrence\cite%
{wooter} which is defined by

\begin{equation}
C(\rho )=\mathrm{max}(0,\text{ }\lambda _{1}-\lambda _{2}-\lambda
_{3}-\lambda _{4})
\end{equation}

where $\lambda _{1},\lambda _{2},\lambda _{3},\lambda _{4}$ is the square
root of non-Hermitian matrix $R$ in decreasing order, and

\begin{equation}
R=\rho \left( \sigma _{1}^{y}\otimes \sigma _{1}^{y}\right) \rho ^{\ast
}\left( \sigma _{1}^{y}\otimes \sigma _{1}^{y}\right) .
\end{equation}

We consider that when the initial state of junctions is $\left\vert
0\right\rangle _{1}\left\vert 0\right\rangle _{2}$ and the controllable
microwave field is prepared in coherent state. Then, through a simple
calculation, we obtain the concurrence of the states of JJ qubits subsystem
Eq.($33$) is

\begin{equation*}
C=\sqrt{2}\sqrt{\left( \sqrt{AB}-\left\vert D\right\vert \right) ^{2}}
\end{equation*}%
\begin{eqnarray}
A &=&e^{-\left\vert \alpha \right\vert ^{2}}\underset{k}{\sum }\dfrac{%
\left\vert \alpha \right\vert ^{2k}}{k!k!}\left\vert \cos \left( \omega
_{k}t\right) -i\sin \left( \omega _{k}t\right) \sin \left( \theta
_{k}\right) \right\vert ^{2}  \notag \\
B &=&e^{-\left\vert \alpha \right\vert ^{2}}\underset{k}{\sum }\dfrac{%
\left\vert \alpha \right\vert ^{2k}}{k!k!}\sin ^{2}\left( \omega
_{k}t\right) \cos ^{2}\left( \theta _{k}\right)  \notag \\
D &=&e^{-\left\vert \alpha \right\vert ^{2}}\underset{k}{\sum }\dfrac{%
\left\vert \alpha \right\vert ^{2k}}{k!k!}\sin \left( \omega _{k}t\right)
\cos \left( \theta _{k}\right)  \notag \\
&\cdot &\left\{ \cos \left( \theta _{k}\right) \cos \left( \omega
_{k}t\right) -i\sin \left( \omega _{k}t\right) \sin \left( \theta
_{k}\right) \right\} .
\end{eqnarray}%
\begin{figure}[tbp]
\includegraphics[angle=-90,width=7cm]{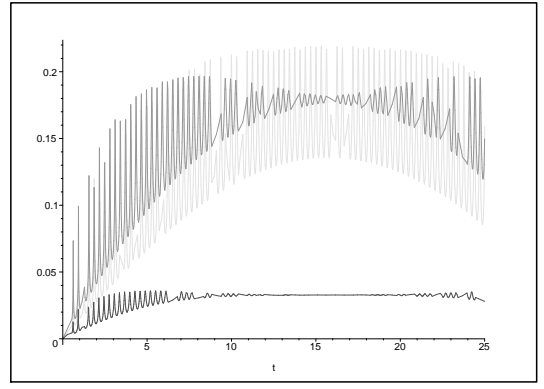}
\caption{ The vertical axis represent concurrence $C$, the horizonal axis
represent time $t$, $\protect\alpha =0.1$ (light ashen line), $\protect%
\alpha =1.1$ (ashen line) and $\protect\alpha =3$ (black line). }
\label{fig:bell-cohen}
\end{figure}
\bigskip

In FIG.6, the concurrences of the qubits are plotted for different values of
$\alpha $. It is seen that, with the increasing the eigenvalue of the
controllable microwave field, the concurrence decrease sharply.

\bigskip We study another case, when the initial state of the controllable
micro-wave field is thermal state

\begin{equation}
\rho _{mw}=\frac{1}{Z}\underset{n}{\sum }e^{-n\beta E}\left\vert
n\right\rangle \left\langle n\right\vert \text{ \ \ }Z=\underset{n}{\sum }%
e^{-n\beta E},
\end{equation}%
the state of the total system evolute into%
\begin{equation}
\rho =\frac{1}{Z}\underset{n}{\sum }e^{-n\beta E}\left\vert \psi
_{n}\right\rangle \left\vert n\right\rangle \left\langle n\right\vert
\left\langle \psi _{n}\right\vert \ .\
\end{equation}%
By a simple calculation, we get the RDM of the JJ qubits

\begin{equation}
\rho _{jj}=\frac{1}{Z}\underset{n}{\sum }e^{-n\beta E}\left\vert \psi
_{n}\right\rangle \left\langle \psi _{n}\right\vert \ .
\end{equation}

\bigskip By the same way, we calculate the concurrence of this state in the
following

\begin{eqnarray}
C &=&\sqrt{2}\sqrt{\left( \sqrt{AB}-\left\vert D\right\vert \right) ^{2}} \\
A &=&\frac{1}{Z}\underset{k}{\sum }e^{-k\beta E}\left\vert \cos \left(
\omega _{k}t\right) -i\sin \left( \omega _{k}t\right) \sin \left( \theta
_{k}\right) \right\vert ^{2}  \notag \\
B &=&\frac{1}{Z}\underset{k}{\sum }e^{-k\beta E}\sin ^{2}\left( \omega
_{k}t\right) \cos ^{2}\left( \theta _{k}\right)  \notag \\
D &=&\frac{1}{Z}\underset{k}{\sum }e^{-k\beta E}\sin \left( \omega
_{k}t\right) \sin \left( \theta _{k}\right)  \notag \\
&\cdot &\left\{ \cos \left( \theta _{k}\right) \cos \left( \omega
_{k}t\right) -i\sin \left( \omega _{k}t\right) \sin \left( \theta
_{k}\right) \right\}  \notag
\end{eqnarray}%
\begin{figure}[tbp]
\includegraphics[angle=-90,width=7cm]{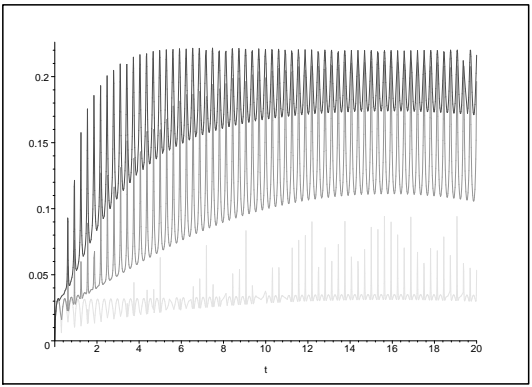}
\caption{ The vertical axis represent concurrence $C$, the horizonal axis
represent time $t$, with different parameter$\protect\beta E$ of the thermal
states. $\protect\beta E=0.7$ (light ashen line),$\protect\beta E=2$ (ashen
line) and $\protect\beta E=6$ (black line) }
\label{fig:bell-cohen}
\end{figure}
\bigskip In FIG.7, the concurrence of the state $\rho _{jj}$ is periodic
function of time $t$ and the concurrence $C\rightarrow $ the maximal value,
when $\beta E\rightarrow \infty $. This is because $\beta E\rightarrow
\infty $, the thermal state $\rho _{mw}\rightarrow \left\vert 0\right\rangle
\left\langle 0\right\vert $. \bigskip

FIG.6 and FIG.7 illustrate that any types of the controllable micro-wave
field can not increase the entanglement of the JJ qubits when its initial
state is superposition of $\left\vert 00\right\rangle $ and $\left\vert
11\right\rangle $, and the maximal entanglement is much smaller than $1$.
Only the initial state of the controllable micro-wave field is vacuum state,
the entanglement can reach the maximal value.

\bigskip

\section{Conclusion}

In summary, we study the protocols which can create maximally entangled
states between two qubit coupled to a controllable microwave field in a
cavity. In order to obtain the analytic study for this decoherence problem,
we generalized Fr\H{o}hlich transformation to re-derive the effective
Hamiltonian of these system, which is equivalent to that obtained from the
adriatic elimination approach. Because of nontrivial decoherence, we can not
construct an ideal logic gate by this system. But we can construct a
decoherence-free subspace of two-dimension to against this adiabatic
decoherence in this system. \ \

\section{acknowledgment}

We thank prof. C.P. Sun for helpful discussions.This work was
partially supported by the CNSF (grant No.90203018) ,the Knowledge
Innovation Program (KIP) of the Chinese Academy of Sciences , the
National Fundamental Research Program of China with
No.001GB309310, K. C. Wong Education Foundation, HongKong, and
China Postdoctoral Science Foundation.

\end{document}